\documentclass[aps,prd,reprint,groupedaddres]{revtex4-2}

\usepackage[
colorlinks=true,
filecolor=black,
anchorcolor=blue,
linkcolor=blue,
citecolor=cyan,
urlcolor=cyan,
linktocpage=true,
plainpages=false,
breaklinks=true,
pdfstartview=FitH
]{hyperref}

\usepackage[utf8x]{inputenc}
\DeclareUnicodeCharacter{2212}{\textendash}
\usepackage{graphicx}
\usepackage{amssymb}
\usepackage{amsmath}
\usepackage{amsthm}

\usepackage{hyperref}
\usepackage{multirow}
\usepackage{bigints}
\usepackage{subfigure}
\usepackage{dcolumn}
\usepackage{bm}
\usepackage{color}
\hypersetup{colorlinks=true,linkcolor=blue,citecolor=cyan,urlcolor=cyan,bookmarks=true} 

\usepackage{chngcntr}
\counterwithin{equation}{section}

\begin{document}

\allowdisplaybreaks

\title{Artificial Precision Polarization Array:\\
Sensitivity for the axion-like dark matter with clock satellites}

\author{Hanyu Jiang$^{1,3}$}
\email[]{jianghy@bao.ac.cn}

\author{Baoyu Xu$^{1,3}$}
\email[]{xuby@bao.ac.cn}

\author{Yun-Long Zhang$^{1,2}$}
\email[Corresponding author:~]{zhangyunlong@nao.cas.cn}

\affiliation{$^1$National Astronomical Observatories, Chinese Academy of Sciences,
Beijing 100101, China}

\affiliation{$^2$School of Fundamental Physics and Mathematical Sciences, Hangzhou Institute for Advanced Study, University of Chinese Academy of Sciences, Hangzhou 310024, China}

\affiliation{$^3$School of Astronomy and Space Science, University of Chinese Academy of Sciences, Beijing 100049, China
}
 
\begin{abstract} 
The approaches to searching for axion-like signals based on pulsars include observations with pulsar timing arrays (PTAs) and pulsar polarization arrays (PPAs). However, these methods are limited by observational uncertainties arising from multiple unknown and periodic physical effects, which substantially complicate subsequent data analysis. To mitigate these issues and improve data fidelity, we propose the Artificial Pulsar Polarization Arrays (APPA): a satellite network comprising multiple pulsed signal transmitters and a dedicated receiver satellite. To constrain the axion-photon coupling parameter $g_{a\gamma}$, we generate simulated observations using Monte Carlo methods and investigate the sensitivity of APPA using two complementary approaches: Likelihood analysis and frequentist analysis. Simulations indicate that for the axion mass range of $10^{-22}-10^{-18}$ eV, APPA yields a tighter upper limit on $g_{a\gamma}$ (at the 95\% C.L.) than conventional ground-based observations, while also achieving superior detection sensitivity. Moreover, a larger spatial distribution scale of the satellite network corresponds to a greater advantage in detecting axions with lighter masses.
\end{abstract}

\maketitle
\tableofcontents

\section{Introduction}\label{Sec:1}
The strong CP problem in quantum chromodynamics was famously addressed by Peccei and Quinn, who postulated a new $U_{PQ}(1)$ symmetry \cite{10.1103/PhysRevLett.38.1440, 10.1103/PhysRevD.16.1791}. Now we know that the $U_{PQ}(1)$ breaking yields a pseudo Nambu–Goldstone boson: the axion \cite{10.1103/PhysRevLett.40.223, 10.1103/PhysRevLett.40.279,10.1088/1126-6708/2006/06/051,10.1007/BF01570798}. The axion can also be identified as a light particle whose couplings could make it a viable dark matter (DM) candidate \cite{10.1016/0370-2693(83)90637-8,10.1016/0370-2693(83)90639-1,10.1016/0370-2693(83)90638-X}. The ultralight ALPs (with masses $m_a\ll10^{-20}$eV) share the property that their de Broglie wavelength in galaxies is astrophysically large, implying a large occupation number in galactic halos and a collective wave like behavior \cite{2507.00705,10.1146/annurev-astro-120920-010024,10.1088/1475-7516/2014/02/019}. It was recognized that such fuzzy DM \cite{10.1103/physrevlett.85.1158} could address apparent discrepancies of the collisionless cold DM paradigm on small scales. In particular, it was proposed that a boson of mass $m_a\sim\mathcal{O}\big(10^{-22}\big)$eV would stabilize gravitational collapse below  kpc scales, yielding diffuse halo cores and strongly suppressing subgalactic power \cite{10.1103/physrevlett.85.1158}. These wave effects, originating from the effective ``quantum pressure" of the scalar field, modify the linear matter power spectrum and produce solitonic cores at halo centers.

Cosmological observations provide important and effective methods of constraining ultralight dark matter (ULDM) \cite{10.1103/PhysRevD.95.043541, 10.1007/s00159-021-00135-6, 2203.14915}. Analyses of the Planck/WMAP data indicate that axions with $m_a\lesssim10^{-24}$eV may constitute a fraction of the dark matter density (at 95\% C.L.) \cite{10.1103/PhysRevD.91.103512}. Tracking the small scale matter power spectrum with the Lyman-$\alpha$ forest has significantly improved the sensitivity of axion searches \cite{10.1103/PhysRevLett.119.031302,10.1016/j.physrep.2016.06.005} and has excluded portions of the allowed axion mass range \cite{10.1103/physrevlett.126.071302}. On galactic scales, astrophysical observations also place stringent limits on axions. For example, stellar kinematic analyses of dwarf spheroidal galaxies and low surface brightness halos can test the solitonic cores predicted by fuzzy dark matter \cite{10.1038/nphys2996, 10.1103/PhysRevLett.113.261302, 10.1093/mnras/stx1887, 10.1103/physrevlett.123.051103, 10.3847/2041-8213/abf501, 10.1051/0004-6361/202346686}; jeans analyses of observational data from different galaxies further constrain the mass range of axions and provide estimates of the core radii of dark matter halos \cite{10.1093/MNRAS/STX1941}. Additional approaches for probing ULDM include searches for ULDM induced central cores using rotation curves and stellar streams \cite{10.1103/physrevd.105.083015, 10.1051/0004-6361/202346686, 1808.00464, 2405.19410, 10.1088/1475-7516/2025/06/050}, as well as constraints on the upper limit of the axion density derived from eROSITA galaxy cluster counts and weak lensing data \cite{2502.03353}.

Novel timing and polarization techniques provide additional search strategies.
In \cite{10.1088/1475-7516/2014/02/019}, it was noted that an oscillating ULDM field would induce a characteristic, nearly monochromatic perturbation in pulsar timing residuals. PPTA \cite{10.1103/PhysRevD.98.102002}, NANOGrav \cite{10.1088/1475-7516/2020/09/036}, and EPTA \cite{10.1103/PhysRevD.111.062005} pulsar timing data have been used to look for this signature: the data lead to upper limits on the axion coupling and local density in the mass range $m_{a}\sim\mathcal{O}\big(10^{-23}–10^{-21}\big)$eV. Independently, ultralight axions coupled to photons can cause cosmic birefringence: coherent ALP oscillations rotate the plane of linear polarization of passing light. Constraints on the axion–photon coupling from birefringence effects in polarized millisecond-pulsar pulses and polarized fast radio bursts (FRBs) are given in \cite{10.1103/PhysRevD.111.062005} and \cite{10.1038/s42005-025-02045-w}.
In parallel, the mechanism of arraying many polarized sources has emerged: in \cite{10.1103/physrevlett.130.121401}, the idea of a “pulsar polarization array(PPA)” was introduced to statistically enhance sensitivity to axion birefringence. For further advances on constraining axion parameters through polarization measurements of signals from black holes, please see \cite{10.1103/PhysRevD.110.063039}.

In this work, we explore a novel avenue to probe this frontier using next generation timing and polarization networks. In particular, the concept of Artificial Precision Timing Array (APTA) was recently proposed in \cite{2401.13668}, where a constellation of clock satellites acts as artificial pulsars, to detect decihertz gravitational waves. The APTA detects subtle variations in signal parameters that may be induced by decihertz gravitational waves through precise and controllable signals. This enables the identification of the presence of decihertz gravitational waves. Compared with naturally occurring astrophysical PTAs, the APTA is far less affected by environmental and astrophysical uncertainties, providing a significant advantage. By analogy, one can envision the Artificial Precision Polarization Array (APPA) of precisely calibrated sources to detect axion induced birefringence. The APPA can detects the possible existence of axion induced birefringence by transmitting precisely controlled pulse signals with well defined polarization parameters. In this work, we investigate how the APPA can be leveraged to constrain or discover ultralight axion DM, focusing on the unique parameter ranges and systematics accessible to such terrestrial–space hybrid experiments.

This paper is organized as follows. In Sec.~\ref{Sec:2}, we provide a brief overview of the axion induced birefringence effect and the principle of detecting axion signals through the rotation of the linear polarization angle. In Sec.~\ref{Sec:3}, we introduce the APPA concept in detail, likelihood analysis as described in Sec.~\ref{subSecLikelihood} and the Generalized Lomb–Scargle Periodogram (GLSP) as presented in Sec.~\ref{subSecFre}. These methods are employed to derive the 95\% C.L. upper limit $g_{95\%}$ and to compute sensitivity curves for the axion–photon coupling parameter $g_{a\gamma}$. The conclusions are summarized in Sec.~\ref{Sec:4}. Throughout this work, we adopt natural units with $c=\hbar=1$ and choose the metric signature as $(-,+,+,+)$.

\begin{figure}[!htb]
    \centering
    \includegraphics[width=0.45\textwidth]{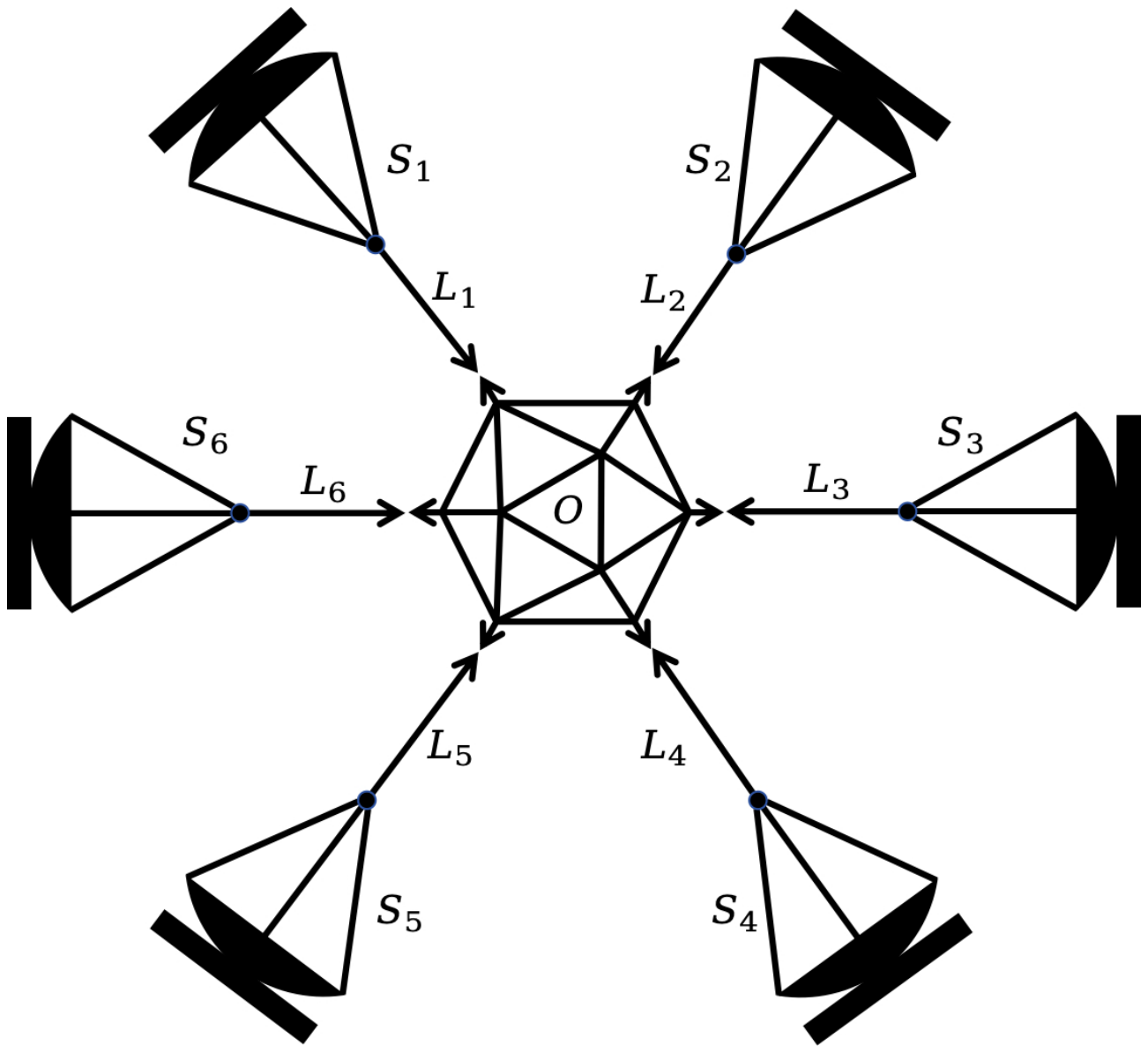}
    \caption{Conceptual diagram for APPA. The satellite network consists of a single receiving satellite $O$ and several transmitting satellites $S_{i} (i=1,...,6)$, each equipped with an ultra precision clock. This diagram is purely illustrative and does not represent the actual physical configuration.}
    \label{satellite}
\end{figure}

\section{Axion induced birefringence effect}\label{Sec:2}

In this section, we present the axion induced birefringence and show how it produces a rotation of the polarization plane of propagating electromagnetic waves. We then describe how measurements of this polarization rotation can be employed to constrain axion related parameters, in particular the axion–photon coupling $g_{a\gamma}$ and the axion mass $m_{a}$.

The action can be written as $S=\int d^{4}x\sqrt{-g}\mathcal{L}$, where $\mathcal{L}$ denotes the Lagrangian density:
    \begin{equation}
        \begin{aligned}
 \mathcal{L}&=-\frac{1}{4}F_{\mu\nu}F^{\mu\nu}-\frac{g_{a\gamma}}{4}aF_{\mu\nu}\widetilde{F}^{\mu \nu}-\frac{1}{2}\partial^{\mu}a\partial_{\mu}a-\frac{1}{2}m^{2}_{a}a^{2}.
        \end{aligned}
        \label{SL}
    \end{equation}
The axion field $a$ with mass $m_{a}$ can couple to the electromagnetic field $F_{\mu\nu}=\partial_{\mu}A_{\nu}-\partial_{\nu}A_{\mu}$, and $\widetilde{F}^{\mu \nu}\equiv \frac{1}{2}\varepsilon^{\mu\nu\rho\sigma}F_{\rho\sigma}$ represents the dual electromagnetic tensor, and $g_{a\gamma}$ is the axion coupling coefficient.

Starting from the Lagrangian density $\mathcal{L}$, and by applying  Euler–Lagrange equation, we obtain the modified Maxwell equations  that incorporate axion induced terms,
    \begin{equation}
        \partial_{\mu}F^{\mu\nu }=g_{a\gamma}\partial_{\mu}a\widetilde{F}^{\mu \nu}.
        \label{EMequ}
    \end{equation}
The birefringence effect for the electromagnetic field induced by axion field can be expressed as \cite{10.1103/PhysRevD.111.062005,10.1088/1475-7516/2022/06/014},
    \begin{equation}
        \Delta \omega =\pm \frac{1}{2} g_{a\gamma }\left(\partial_{t}a+\nabla a\cdot \hat{k}\right).
        \label{deltaomega}
    \end{equation}
Here $\hat{k}$ denotes the unit wave vector, and the details of the derivation are given in Appendix \ref{App:A}. In an axion background, electromagnetic propagation is dispersive: the left and right circularly polarized modes acquire a tiny frequency splitting. Consequently, axion induced birefringence produces a relative phase shift between the two chiral polarization components \cite{10.1103/PhysRevD.111.062005,10.1088/1475-7516/2022/06/014,2412.02229},
    \begin{equation}
        \Delta\phi=\frac{g_{a\gamma}}{2}\int_{t_{s}}^{t_{o}}\frac{da}{dt}dt=\frac{g_{a\gamma}}{2}\left(a_{o}-a_{s}\right).
        \label{deltaphi}
    \end{equation}
Here, $a_{s}$ and $a_{o}$ denote the axion field at the signal emission point and the signal reception point, respectively.
Unlike Faraday rotation, which depends on the observing frequency and on the line of sight magnetic field encountered along the propagation path, the polarization plane rotation induced by axion birefringence is determined by the axion field values at the endpoints. Consequently, the axion induced rotation is monochromatic.

In \cite{10.1088/1475-7516/2014/02/019}, the authors considered the occupation number of a scalar axion field and estimated it under standard local halo assumptions. Assuming a characteristic dark-matter velocity $v\sim10^{-3}c$,
and a local dark matter energy density $\rho_{\text{DM}}=0.4$GeV/cm$^{3}$ in the vicinity of the Solar System, the occupation number can be estimated by
    \begin{equation}
        \frac{\rho_{\text{DM}}}{m\cdot \left(mv\right)^{3}}\simeq 10^{95}\left(\frac{\rho_{\text{DM}}}{0.4\mathrm{GeV/cm}^{3}}\right)\left(\frac{10^{-23}\mathrm{eV}}{m}\right)^{4}\left(\frac{10^{-3}}{v}\right)^{3}.
        \label{ocunum}
    \end{equation}
Such a large occupation number suppresses quantum fluctuations of the axion field, which enabling it to be treated as a classical field and thereby simplifying the analysis,
    \begin{equation}
        a\big(\vec{x}, t\big) =a_{0}\big(\vec{x}\big)\mathrm{cos}\big[m_{a}t+\varphi\big(\vec{x}\big)\big].
        \label{axion}
    \end{equation}
The amplitude of the oscillating scalar axion field is determined by the local dark matter energy density~\cite{10.1088/1475-7516/2019/02/059}:
    \begin{equation}
        \rho_{\text{DM}}=\frac{1}{2}m_{a}^{2}\left \langle a^{2} \right \rangle.
        \label{rouDM}
    \end{equation}
Meanwhile, with the typical velocity $v\sim10^{-3}c$, we compile a list of useful physical quantities, including the Compton frequency $f_{\mathrm{c}}$, the de Broglie wavelength $\lambda_{\mathrm{de}}$ and the coherence time $\tau_{\mathrm{coh}}$,
    \begin{equation}
        \begin{aligned}
 f_{\mathrm{c}}&=\frac{m_{a}}{2\pi}\approx 2.42\times  10^{-8}\mathrm{Hz} \Big(\frac{m_{a}}{10^{-22}\mathrm{eV}}\Big),\\
\lambda_{\mathrm{de}}&=\frac{2\pi}{m_{a}v}\approx 4.02\times 10^{2}\mathrm{pc} \Big(\frac{10^{-22}\mathrm{eV}}{m_{a}}\Big),\\
\tau_{\mathrm{coh}}&=\frac{1}{m_{a}v^{2}}\approx 2.09\times 10^{5}\mathrm{yr} \Big(\frac{10^{-22}\mathrm{eV}}{m_{a}}\Big).
        \end{aligned}
        \label{fdt}
    \end{equation}
For a given axion mass, if the condition $\tau_{\mathrm{coh}} \gg T_{\text{obs}}$ is satisfied, then the observed axion field amplitude behaves stochastically.

As shown in \cite{10.1103/physrevd.97.123006, 10.1038/s41467-021-27632-7,10.1088/1475-7516/2022/06/014}, within a single coherence domain the amplitude distribution of the axion field follows a Rayleigh distribution,
    \begin{equation}
a_{0}=\frac{\sqrt{2\rho_{\text{DM}}}}{m_{a}}\alpha,\qquad 
p\big(\alpha\big)\sim \alpha e^{-\frac{\alpha^{2}}{2}}. 
        \label{relydis}
    \end{equation}
By combining Eq.~(\ref{deltaphi}), Eq.~(\ref{rouDM}) and Eq.~(\ref{relydis}), we obtain the expression for the general case
    \begin{equation}
        \Delta\phi=\phi_{a}\mathrm{cos}\big(m_{a}t+\varphi_{a}\big),
        \label{deltaphiphia}
    \end{equation}
where
    \begin{equation}
        \begin{aligned}
            \phi_{a}&=\frac{g_{a\gamma}}{\sqrt{2}m_{a}}\left(\rho_{o}\alpha_{o}^{2}+\rho_{s}\alpha_{s}^{2}-2\sqrt{\rho_{o}\rho_{s}}\alpha_{o}\alpha_{s}\mathrm{cos}\theta_T \right)^{\frac{1}{2}},\\
            \varphi_{a}&=\arctan\frac{\sqrt{\rho_{o}}\alpha_{o}\sin\varphi_{o}+\sqrt{\rho_{s}}\alpha_{s}\sin\big(m_{a}T-\varphi_{s}\big)}{\sqrt{\rho_{o}}\alpha_{o}\cos\varphi_{o}-\sqrt{\rho_{s}}\alpha_{s}\cos\big (m_{a}T-\varphi_{s}\big)},
        \end{aligned}
        \label{phiadelta}
    \end{equation}
with $\theta_T\equiv m_{a}T+\varphi_{o}-\varphi_{s}$, and $T$ is the signal propagation time.\color{black}

\section{Methods of analysis}\label{Sec:3}
In previous studies, potential axion signals have typically been investigated through long-term monitoring with pulsar timing arrays (PTAs) and pulsar polarization arrays (PPAs). However, natural pulsar signals are subject to several limitations:

{\it 1. Wide distance distribution of pulsars.} The known pulsar population is predominantly concentrated within distances of a few kpc \cite{10.1051/0004-6361/202243305}. The nearest pulsars lie at 120–170 pc, such as PSR J0437−4715 at about 157 pc \cite{10.1086/529576} and PSR J2144−3933 at about 170 pc \cite{10.1111/j.1745-3933.2011.01009.x}. Over such large propagation distances, pulsar signals are affected by a variety of complex astrophysical processes that tend to obscure potential axion induced signatures. These effects collectively complicate the modeling and degrade the achievable signal-to-noise ratio (S/N).

{\it 2. Uncertainty in local dark matter energy density.} As indicated in Eq.~(\ref{phiadelta}), the measurement of the axion–photon coupling parameter $g_{a\gamma}$ requires knowledge of the dark matter energy densities $\rho_{s}$ near the pulsar and $\rho_{o}$ near the Solar System. However, $\rho_{s}$ is difficult to determine observationally and is subject to large uncertainties, which ultimately limit the precision of the derived constraints.

{\it 3. Randomness of phase.} For pulsars, the pulsar and the observer often reside in different axion coherent regions. This means the phase $\varphi_{s}$ of the axion field at a pulsar is inherently random and indeterminate, which prevents the signal propagation time $T$ from being used to reliably determine the value of $\theta_{T}$ in Eq.~(\ref{phiadelta}). Meanwhile, the phase of the pulsar signals cannot be well determined, introducing uncertainty into practical research and analysis.

{\it 4. Complex and frequency-dependent ionospheric interference.} In practical observations, most pulsar signals are detected in the radio band, and ground-based radio telescopes therefore serve as the primary observational instruments. However, such measurements are inevitably affected by ionospheric interference. The Earth’s ionosphere induces Faraday rotation, which modifies all pulsar signals and leaves a characteristic imprint in the data. Consequently, this effect must be corrected during data processing. Although physical models of the Earth’s ionosphere have been extensively developed \cite{10.1051/0004-6361/201220728,10.1093/mnras/sty3324,10.1029/2012RS004992}, and a variety of algorithms for mitigating ionospheric contamination have been proposed \cite{10.1086/501444,10.1051/0004-6361/201220728,10.1007/s00190-008-0266-1}, these corrections nonetheless impose an additional computational burden.

To address these observational challenges at their source, we extend the concept proposed in \cite{2401.13668} by introducing the APPA, a network of space-based satellites equipped with ultra precise clocks and capable of regularly transmitting pulsed signals to a dedicated receiving satellite. The transmitted signals are strictly periodic, with a timing uncertainty of $\sigma_{t}=\xi_{\delta t}\delta t$ \cite{2401.13668}, and carry preconfigured polarization information. As illustrated in Fig.~\ref{satellite}, the receiving satellite, positioned at the center of the transmitting network, continuously collects APPA signals and can accurately measure both the distance to and the direction of each transmitting satellite. We assume that the entire satellite network is deployed within the solar system, therefore it can be treated as residing within a single coherent axion field, implying $\varphi_{o}=\varphi_{s}=\varphi$, $\rho_{o}=\rho_{s}=\rho_{\text{DM}}$ and $\alpha_{o}=\alpha_{s}=\alpha$ in Eq.~(\ref{phiadelta}).
Consequently, a significant fraction of environmental disturbances and intrinsic signal uncertainties are effectively eliminated, and Eq.~(\ref{phiadelta}) reduces to
    \begin{equation}
        \begin{aligned}
            \phi_{a}&=\frac{g_{a\gamma}}{m_{a}}\sqrt{\frac{\rho_{\text{DM}}}{2}}\alpha \big[2\left(1-\mathrm{cos}\left(m_{a}T\right)\right)\big]^{\frac{1}{2}},\\
            \varphi_{a}&=\arctan\frac{\sin\varphi+\sin\big(m_{a}T-\varphi\big)}{\cos\varphi-\cos\big (m_{a}T-\varphi\big)}.
        \end{aligned}
        \label{phiapart3}
    \end{equation}
To demonstrate the detection advantage of APPA and simplify the analysis, our work is based on the following assumptions: ($i$) the satellite network remains confined to a single coherent axion domain; ($ii$) we consider only the scenario in which all satellites remain stationary relative to one another; ($iii$) the random noise $\lambda$ is limited to Gaussian white noise,
    \begin{equation}         
        \begin{aligned}
            \lambda=\big(EF\sigma_{\text{err}}\big)^{2}+EQ^{2}.
        \end{aligned}     
        \label{lamda}     
    \end{equation}
Here, $\sigma_{\text{err}}$ denotes the instrumental measurement error, which we take to be $\sigma_{\text{err}}=0.3^{\circ}$. Eq.~(\ref{lamda}) accounts for measurement uncertainties, which are modeled by two parameters $EF$ and $EQ$, with $EF$ being the white noise scaling factor and $EQ$ being the additional white noise. Both of them are free parameters. For more details on white noise, please refer to \cite{2412.02229,10.1103/PhysRevD.98.102002,10.1093/mnras/staa3411,10.1088/0004-637x/813/1/65,10.1103/PhysRevD.106.L081101,10.1103/PhysRevD.111.062005}. Here, the prior distributions are specified as follows:
    \begin{equation}         
        \begin{aligned}
            P\big(EF\big)&=\mathcal{N}\big[1,1\big],\\
            P\left(\mathrm{log_{10}}(EQ\mathrm{/rad})\right)&=\mathcal{N}\big[-5,1\big].
        \end{aligned}         
        \label{EFEQ}
    \end{equation}

\begin{figure}[!htb]
        \centering
        \includegraphics[width=0.45\textwidth]{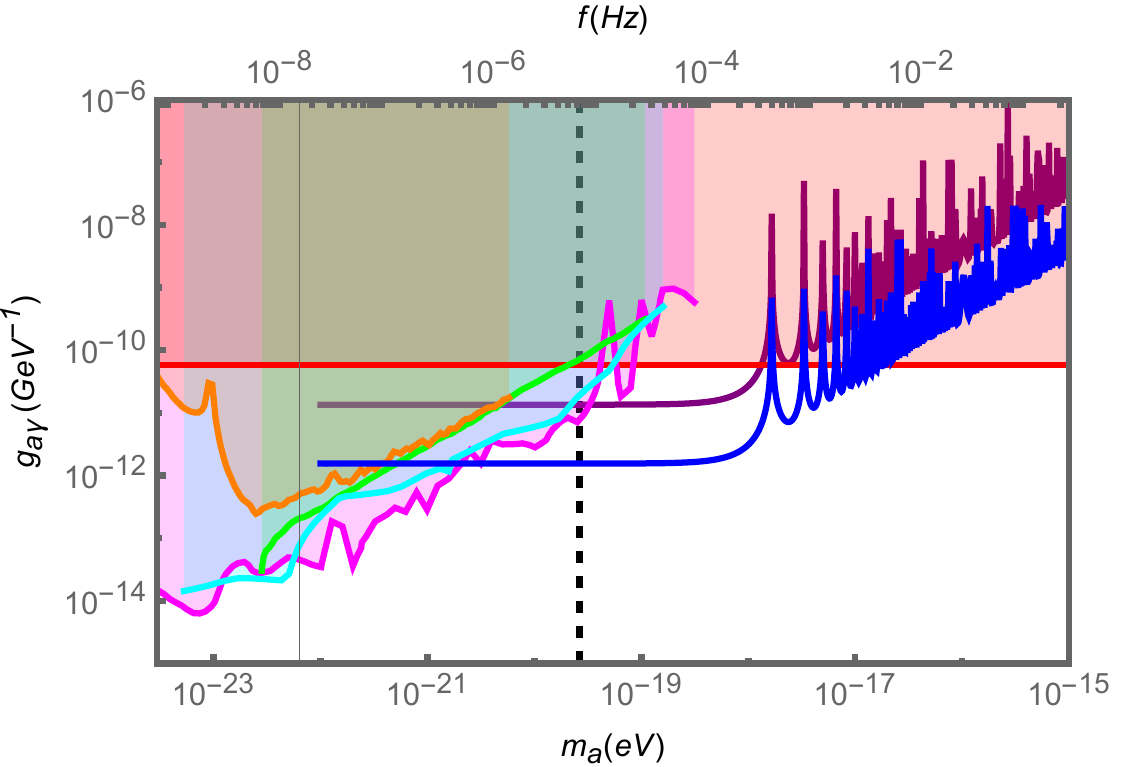}
        \caption{Comparison between the APPA simulation results and existing observational constraints. Color coding: Red, CAST \cite{10.1038/nphys4109}; Magenta, PPA \cite{2412.02229}; Orange, EPTA \cite{10.1103/PhysRevD.111.062005}; Cyan, PSR J0437−4715 \cite{10.1103/PhysRevD.100.063515}; Green, PPTA \cite{10.1088/1475-7516/2022/06/014}; Blue, likelihood analysis ($L=5$AU), we shows that APPA has a detection advantage within $m_a\sim\mathcal{O}\big(10^{-22}–10^{-18}\big)$eV and exhibits comparable sensitivity to axions with different masses; Purple, frequentist analysis, the detection sensitivity is set to $\phi_{a}=0.003^{\circ}$ in Eq.~(\ref{deltaphisim}), consistent with current observational limits on linear polarization angle rotation, with minor fluctuations attributed to instrumental noise. The black dashed line marks the critical boundary, to the left of this boundary, the condition $\mathrm{cos}\big(m_{a}T\big)\simeq1$ holds true. To the right it is no longer valid, we adopt $m_{a}T=0.1$ as the reference criterion.}
        \label{compare}
    \end{figure}

First, we show the main conclusions of this work in advance. We present a comparison between the simulation results of this work and existing observational constraints in a single figure (see Fig.~\ref{compare}). Owing to the long observational baseline $T_{\text{obs}}$ and the highly stable signal emission cadence $\Delta t$, APPA is theoretically capable of probing an exceptionally broad range of axion induced oscillation frequencies. However, as shown in Fig.~\ref{compare}, although APPA theoretically possesses an extremely wide detection range, the domain where it truly demonstrates an advantage in axion signal detection is limited to $m_a\sim\mathcal{O}\big(10^{-22}–10^{-18}\big)$eV, both the 95\% C.L. upper limit $g_{95\%}$ derived from likelihood analysis and the detection sensitivity $g_{a\gamma}$ obtained via frequentist analysis highlight the significant advantages of the APPA framework over current ground-based observations. 

\subsection{Likelihood analysis}\label{subSecLikelihood}
In this section, we employ the parameters of the satellite network defined under the APPA concept and apply the likelihood analysis to compute the 95\% C.L. upper limit of $g_{a\gamma}$.

According to \cite{10.1103/physrevlett.130.121401}, the likelihood function that incorporates the axion signal is given by
    \begin{equation}
        \begin{aligned}
            \mathcal{L}\big(\mathbf{\Theta}|\mathbf{d}\big)=\frac{1}{\sqrt{\mathrm{det}\big(2\pi\mathbf{\Sigma}\big)}}\mathrm{exp}\big(-\frac{1}{2}\mathbf{\widetilde{d}^{T}}\cdot\mathbf{\Sigma}^{-1}\cdot\mathbf{\widetilde{d}}\big),
        \end{aligned}
        \label{likelihood function}
    \end{equation}
where the observational data $\mathbf{d}=\mathbf{s}+\mathbf{n}$ consists of the axion signal $\mathbf{s}$ and noise $\mathbf{n}$, $\mathbf{\widetilde{d}}$ denotes $\mathbf{d}$ with the intrinsic effect subtracted, $\mathbf{\Theta}=\{g_{a\gamma},m_{a}\}$ denotes the set of model parameters and $\mathbf{\Sigma}= \mathbf{\Sigma}^{(s)} +\mathbf{\Sigma}^{(n)}$ is the $N_i\times N_j$ cross-correlation matrix. In practice, each satellite in the network typically generates $N_i$ on the order of $10^{4}\sim10^{5}$ signals. According to \cite{10.1103/physrevlett.130.121401,10.1140/epjc/s10052-011-1554-0,10.1103/PhysRevD.103.076018}, if we given a mass $m_{a}$, the exclusion limit for $g_{a\gamma}$ is set by a test statistic,
    \begin{equation}
        \begin{aligned}
            q\big(g_{a\gamma},m_{a}\big)\equiv2\big[\mathrm{ln}\mathcal{L}\big(\hat{g}_{a\gamma},m_{a}|\mathbf{d}\big)-\mathrm{ln}\mathcal{L}\big(g_{a\gamma},m_{a}|\mathbf{d}\big)\big],
        \end{aligned}
        \label{qdef}
    \end{equation}
where $\hat{g}_{a\gamma}$ maximizes
$\mathcal{L}\big(g_{a\gamma},m_{a}|\mathbf{d}\big)$. With $q(g_{a\gamma},m_{a})=0$ for $g_{a\gamma}<\hat{g}_{a\gamma}$, the upper limit at 95\% C.L. $g_{95\%}$ is given by $q\big(g_{95\%}, m_{a}\big)=2.71$.

As a result, $g_{95\%}$ can be determined from the condition: $\big\langle q \big\rangle =2.71$ \cite{10.1103/physrevlett.130.121401}, where
    \begin{equation}
        \big\langle q\big\rangle=\frac{1}{2}\sum_{i,j}\frac{1}{\lambda_{i}\lambda_{j}}\mathrm{Tr}\left(\mathbf{\Sigma}^{(s)}_{ij}\mathbf{\Sigma}^{(s)}_{ji}\right).
        \label{q}
    \end{equation}
When $i=j$, the trace term in Eq.~\eqref{q} is
    \begin{equation}
        \begin{aligned}
            \mathrm{Tr}\left(\mathbf{\Sigma}^{(s)}_{ii}\mathbf{\Sigma}^{(s)}_{ii}\right)&=4\frac{g_{a\gamma}^{4}}{m_{a}^{4}}N_{i}^{2}\rho_{\text{DM}}^{2}\Big[1-\mathrm{cos}\big(m_{a}T_{i}\big)\mathrm{sinc}\big (y_{ei}\big )\Big]^{2},
        \end{aligned}
        \label{trace1}
    \end{equation}
where $\mathbf{\Sigma}^{(s)}_{ii}$ represents the auto-correlation of the $i$-th source. When $i\ne j$, the trace term in Eq.~\eqref{q} is
    \begin{equation}
        \begin{aligned}
            \mathrm{Tr}\left(\mathbf{\Sigma}^{(s)}_{ij}\mathbf{\Sigma}^{(s)}_{ji}\right)&=\frac{g_{a\gamma}^{4}}{m_{a}^{4}}N_{i}N_{j}\rho_{\text{DM}}^{2}\Big[1+\mathrm{sinc}^{2}\big (y_{ij}\big )\\
            &\quad +2\mathrm{cos}\big(m_{a}\big(T_{i}-T_{j}\big)\big)\mathrm{sinc}\big (y_{ij}\big )\\
            &\quad +f\big(y_{ei},y_{ej}\big)\Big ],
        \end{aligned}
        \label{trace2}
    \end{equation}
where $\mathbf{\Sigma}^{(s)}_{ij}$ represents the cross-correlation between the $i$-th and $j$-th sources.
\color{black}
The factor $f\big(y_{ei},y_{ej}\big)$ in the above expression is given by
    \begin{equation}
        \begin{aligned}
            f\big (y_{ei},y_{ej}\big )&=-2\Big [\mathrm{cos}\big (m_{a}T_{i}\big )\big (\mathrm{sinc}\big (y_{ei}\big )+\mathrm{sinc}\big (y_{ej}\big )\mathrm{sinc}\big (y_{ij}\big )\big )\\
            &\quad +\mathrm{cos}\big (m_{a}T_{j}\big )\big (\mathrm{sinc}\big (y_{ej}\big )+\mathrm{sinc}\big (y_{ei}\big )\mathrm{sinc}\big (y_{ij}\big )\big )\Big ]\\
            &\quad +\Big [\mathrm{sinc}^{2}\big (y_{ei}\big )+\mathrm{sinc}^{2}\big (y_{ej}\big )\\
            &\quad +2\mathrm{cos}\big (m_{a}\big (T_{i}+T_{j}\big )\big )\mathrm{sinc}\big (y_{ei}\big )\mathrm{sinc}\big (y_{ej}\big )\Big ].
        \end{aligned}
        \label{f}
    \end{equation}
Here $T_{i}=L_{i}/c$ is the signal travel time from the $i$-th transmitting satellite to the receiving satellite, and $L_{i}$ is the distance, respectively. $y_{ij}\equiv\left |\mathbf{r}_{i}-\mathbf{r}_{j}\right |/\lambda_{\text{de}}$ and $y_{ei}\equiv\left |\mathbf{r}_{i}\right |/\lambda_{\text{de}}$ are the dimensionless distance parameters and $N_{i}$ denotes the total number of pulse signals emitted by the $i$-th transmitting satellite within the coherence time $\tau_{\mathrm{coh}}$. $\mathrm{sinc}\big(y\big) \equiv\mathrm{sin}\big(y\big)/y$ is the sinc function.
In \cite{10.1103/physrevlett.130.121401}, it has been noted that the factor $f\big(y_{ei},y_{ej}\big)$ is negligible when $y_{ei}\gg 1$, but must be retained when $y_{ei}\le 1$. For pulsars located at cosmological distances, the condition $y_{ei}\gg 1$ is typically satisfied; in contrast, for the APPA considered in this work, the satellite network occupies a single coherent axion domain, the parameter $y_{ei}\le 1$ and in practice $y_{ei}\ll1$, thus $\mathrm{sinc}\big(y_{ei}\big)\simeq1$. Consequently, $f\big(y_{ei},y_{ej}\big)$ plays an important role in the analysis and cannot be neglected.

\begin{figure}[!htb]
        \centering
        \includegraphics[width=0.45\textwidth]{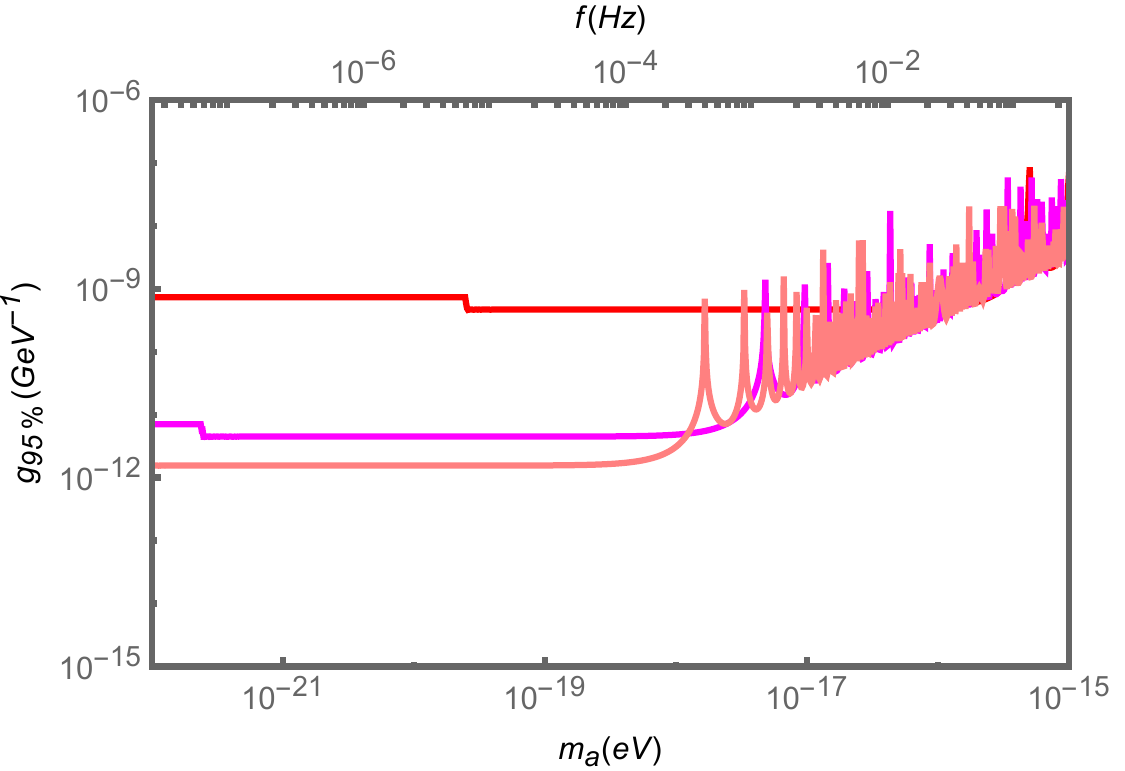}
        \caption{95\% C.L. upper limit $g_{95\%}$ on the function $g_{a\gamma}\big(m_{a}\big)$, which is calculated from $\big\langle q \big\rangle =2.71$ and Eq.~(\ref{q}). Color coding:  Red, $L=2.5\times10^{6}\mathrm{km}$ (LISA \cite{LISA:2017pwj}); Magenta, $L=2.6\times10^{8}\mathrm{km}$ (ASTROD-GW \cite{10.1142/S0218271813410046}); Pink, $L=7.5\times10^{8}\mathrm{km}$ (Jupiter's orbit, 5AU).}
        \label{gaymalikelihood}
    \end{figure}

\begin{figure}[!htb]
        \centering
        \includegraphics[width=0.45\textwidth]{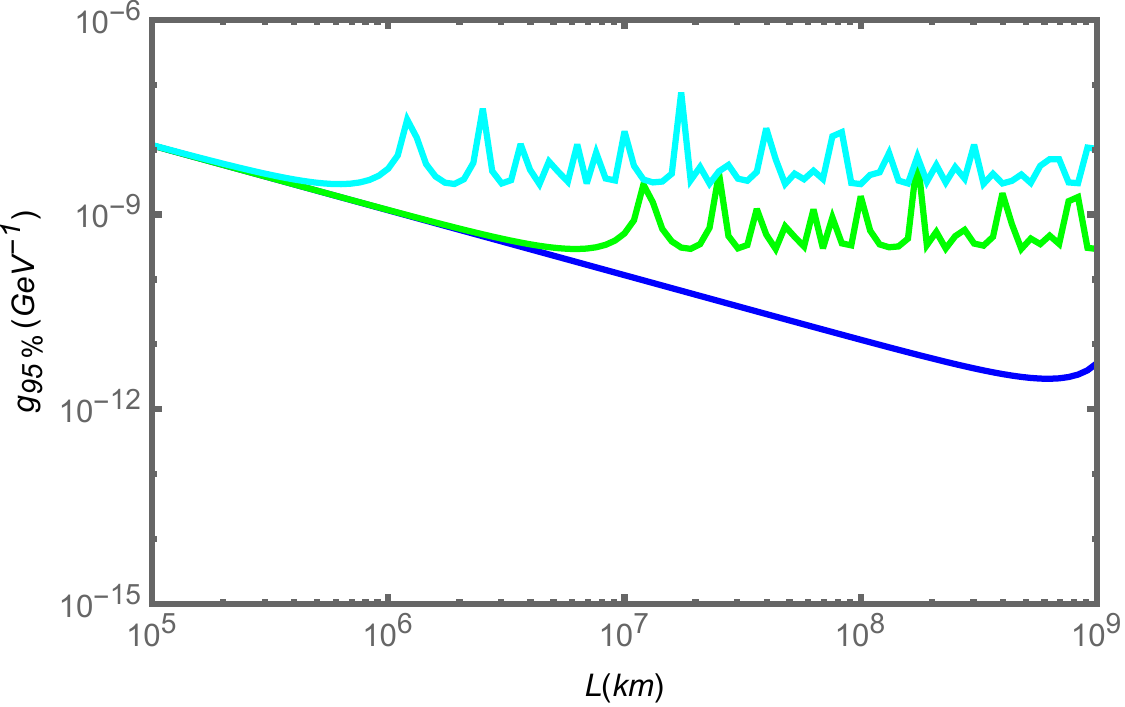}
        \caption{95\% C.L. upper limit $g_{95\%}$ on the function $g_{95\%}\big(L\big)$, which is calculated from $\big\langle q \big\rangle =2.71$ and Eq.~(\ref{q}). The transition point at which $g_{95\%}$ changes from being sensitive to insensitive to the distance is approximately given by the corresponding Compton wavelength $\lambda_{\text{c}}$ of axion. Color coding: Cyan, $m_{a}=10^{-15}\mathrm{eV}$; Green, $m_{a}=10^{-16}\mathrm{eV}$; Blue, $m_{a}=10^{-18}\mathrm{eV}$.}
        \label{gayL}
    \end{figure}

In the numerical simulation, we assume a total of $N_{\mathrm{sat}}=6$ transmitting satellites. The receiving satellite is placed at the center, while the transmitting satellites are randomly distributed in all directions with their distances $L_{i}$. White noise is generated according to Eq.~(\ref{lamda}) and Eq.~(\ref{EFEQ}), with $\sigma_{\text{err}}=0.3^{\circ}$. For each transmitting satellite ($i=1,2,\dots,6$), the number of pulse signals is set to $N_{i}=10^{5}$. Here, using Eq.~(\ref{q}) we show the impact of the axion mass $m_{a}$ and the characteristic scale $L$ (the average distance from the transmitting satellite to the receiving satellite) of the satellite network on the 95\% C.L. upper limits $g_{95\%}$, as illustrated in Fig.~\ref{gaymalikelihood} and Fig.~\ref{gayL}.

Fig.~\ref{gaymalikelihood} indicates that a larger spatial scale of the satellite network is more favorable for detecting axion signals, providing better sensitivity to light axions. Unlike observations using natural pulsars, in our work Eq.~(\ref{q}) incorporates the effect of $f$, for light axions with Compton wavelengths $\lambda_{\text{c}}\ge L$, $g_{95\%}$ shows no significant variation on the order of magnitude. A similar behavior can also be seen in Fig.~\ref{gayL}, which shows the 95\% C.L. upper limits $g_{95\%}$ on the function $g_{a\gamma}\big(L\big)$ for different $m_{a}$. In this figure, the characteristic scale $L$ of the satellite network ranges from $10^{5}\mathrm{km}$ to $10^{9}\mathrm{km}$, covering scales from the Earth–Moon distance to the Jupiter–Sun distance. We find that for lighter axions, the 95\% C.L. upper limit is more sensitive to $L$. A larger $L$ leads to a stronger constraint on $g_{a\gamma}$. Likewise, a saturation scale exists: once $L$ exceeds the Compton wavelength $\lambda_{\text{c}}$ of axion, further increasing does not improve the 95\% C.L. upper limit.

\subsection{Frequentist analysis-GLSP}\label{subSecFre}
In this section, we employ simulated observational data that incorporates scalar axion field signals and analyse the detection sensitivity of APPA based on the Generalized Lomb–Scargle Periodogram (GLSP) method. Our research is based on the following assumptions: the cross-correlation between transmitting satellites can be safely neglected. For further details of the analysis, please refer to Appendix~\ref{App:B}.

The GLSP is a spectral analysis technique designed to detect periodic signals in irregularly sampled time series. It employs a least squares fitting procedure to scan the trial frequencies and compute the power spectrum at each frequency. Compared to traditional Fourier transform methods, the GLSP is particularly advantageous for handling unevenly sampled data, making it well suited for astronomical observations \cite{10.3847/1538-4365/aab766}. According to \cite{10.3847/1538-4365/aab766}, the GLSP is expressed as
    \begin{equation}         
        \begin{aligned}             
            P_{LS}\big(f\big) &= \frac{1}{\sigma^2_y \sum_i \omega_i} \Bigg\{\frac{\big[\sum_i \omega_i\big(y_i - \bar{y}\big)\cos\big(2\pi f\big(t_i - \tau\big)\big)\big]^2}{\sum_i \omega_i \cos^2\big[2\pi f \big(t_i - \tau\big)\big]}\\
            &\quad + \frac{\big[\sum_i \omega_i \big(y_i - \bar{y}\big) \sin\big(2\pi f\big(t_i - \tau\big)\big)\big]^2}{\sum_i \omega_i \sin^2\big[2\pi f\big(t_i - \tau\big)\big]}\Bigg\},
        \end{aligned}         
        \label{PLS}     
    \end{equation}
where
    \begin{equation}         
        \begin{aligned}             
            \tau=\frac{1}{4\pi f}\arctan\frac{\sum_{i}\omega_{i}\sin\big(4\pi ft_{i}\big)}{\sum _{i}\omega_{i}\cos\big(4\pi ft_{i}\big)}, 
        \end{aligned}         
        \label{tao}     
    \end{equation}
and
    \begin{equation}         
        \begin{aligned}
            \omega_{i}&=\frac{1}{\sigma^{2}_{i}},\quad 
            \bar{y}=\frac{\sum_{i}\omega_{i}y_{i}}{\sum_{i}\omega_{i}},\\
            \sigma^{2}_{y}&=\frac{\sum_{i}\omega_{i}\big(y_{i}-\bar{y}\big)^{2}}{\sum_{i}\omega_{i}}.
        \end{aligned}         
        \label{parameter}     
    \end{equation}
Here, $\left\{\sigma_{i}\right\}$ represents instrumental measurement uncertainties, $\left\{\omega_{i}\right\}$ denotes the weights assigned to the data points, while $\left\{t_{i}\right\}$ and $\left\{y_{i}\right\}$ correspond to the observation times and the associated measured data sequence, respectively.
It is emphasized that $\left\{t_{i}\right\}$, $\left\{y_{i}\right\}$ and $\left\{\sigma_{i} \right\}$ are strictly matched in chronological order according to the observation epochs, ensuring a one to one correspondence between each observation and its uncertainty.

A distinctive feature of the window function in ground-based radio telescope observations is that, in order to detect faint pulsar signals, observations are restricted to nighttime. Consequently, the resulting time series is highly non-uniformity sampled. Moreover, due to the annual modulation introduced by the Earth's orbital motion, the data are not only influenced by the ionospheric interference discussed earlier but also subject to additional deterministic periodic effects \cite{10.3847/1538-4365/aab766}. Together, these factors give rise to pronounced sidelobes and spectral aliasing in the Lomb–Scargle periodogram \cite{10.3847/1538-4365/aab766}, thereby increasing the False Alarm Probability (FAP) and potentially obscuring the true signal peaks. As shown in Fig.4 of \cite{10.1088/1475-7516/2022/06/014}, the PPTA observed four representative pulsars (J0437–4715, J1643–1224, J1730–2304 and Crab–QUIJOTE) and presented the GLSP of the time series of polarization measurements. Even after applying the IRI ionospheric corrections, the GLSPs of the four pulsars remain highly irregular. The irregular temporal sampling causes the target axion signal to be masked by numerous spurious peaks in the power spectrum. This effect becomes increasingly severe as the observational S/N decreases, and in extreme cases, the signal peak may fall entirely below the significance threshold, rendering it undetectable. The figure also demonstrates that many interference peaks approach the 32\% FAP level, some of them even approach the 5\% FAP level, substantially complicating the reliable identification of genuine signals.

\begin{figure}[!htb]
        \centering
        \includegraphics[width=0.45\textwidth]{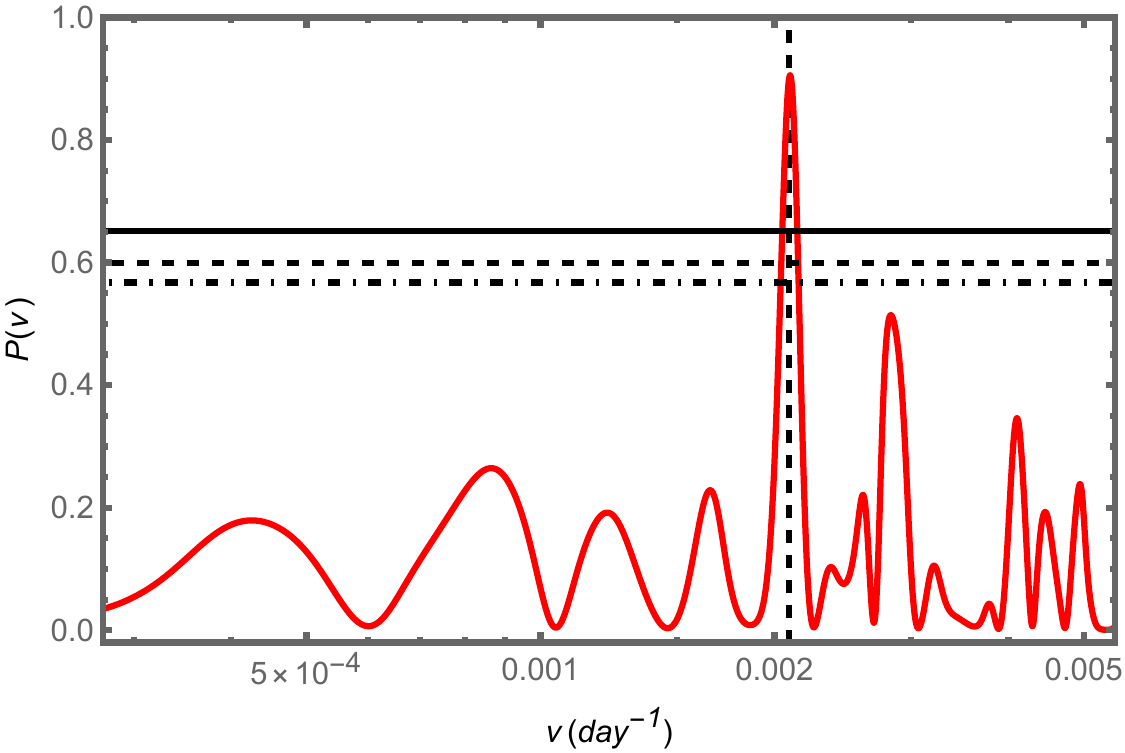}
        \caption{Numerical simulation: The GLSP (red) derived from APPA observational data. The simulated signal includes only Gaussian white noise with S/N=1. The vertical dashed line marks the target signal frequency of the axion with mass $m_{a}=10^{-22} \mathrm{eV}$. The horizontal lines indicate FAP levels of 10\% (dotdashed), 5\% (dashed) and 1\% (solid).}
        \label{regulartimeGLPS}
    \end{figure}

In contrast, within our proposed framework, once the satellite network is deployed, it can operate in continuous monitoring mode. In the idealized case, the corresponding window function of space-based observations reduces to $W_{{t_{i}}} \big(t\big)=1$, thereby eliminating the non-uniformity inherent in ground-based sampling. To illustrate this, we consider a simplified scenario in which the observed signal contains only a single axion induced frequency component, $m_{a}=10^{-22}\mathrm{eV}$. The transmitting satellites emit periodic pulse signals towards the receiving satellite at a fixed interval $\Delta t=\delta t+\sigma_{t}=\big(1+\xi_{\delta t}\big)\delta t$. In addition, the system is affected by Gaussian white noise, $\phi_{\text{noise}}$. Under these assumptions, the APPA measured at the receiving end can be expressed as
    \begin{equation}         
        \begin{aligned}
            \Delta\phi_{\text{sim}}=\phi_{a}\mathrm{cos}\big(2\pi\nu t+\varphi\big)+\phi_{\text{noise}}.
        \end{aligned}
        \label{deltaphisim}     
    \end{equation}
Similar to Fig.4 in \cite{10.1088/1475-7516/2022/06/014}, we consider the case of $\text{S/N}=1$ and a target signal frequency of the axion with mass $m_{a}=10^{-22} \mathrm{eV}$. However, in this scenario, the observation epochs are distributed with high regularity over a continuous 10-year period. The corresponding simulation results are presented in Fig.~\ref{regulartimeGLPS}.
By comparing the simulation results, it is evident that regular observations of pulsed signals substantially enhance the prominence of the target signal peak while simultaneously suppressing spurious interference peaks. This improvement is quantitatively reflected in the lower FAP levels, which provide a clearer statistical distinction between genuine signals and noise.

If the observational data are of low quality and the target frequency peak is completely masked by interference peaks, the GLSP method becomes ineffective. In such cases, for further information, the reader is referred to \cite{10.1103/PhysRevD.111.062005, 10.1088/1475-7516/2022/06/014}. Once the GLSP method successfully extracts the target axion signal frequency $m_{a}$ from the pulsed signals received by the satellite, the APPA sensitivity curve for axion signal detection can be derived using Eq.~(\ref{phiapart3}) and Eq.~(\ref{deltaphisim}).

We assume that the satellite network possesses a detection sensitivity of $0.003^{\circ}$ for linear polarization angle rotation, and that the instrumental noise is parameterized according to Eq.~(\ref{lamda}) and Eq.~(\ref{EFEQ}). Under these assumptions, the sensitivity curve can be obtained subject to the constraint of a total observation time of $T_{\text{obs}}\sim\mathcal{O} \big(10^{1}\big)$ yr with Eq.~(\ref{phiapart3}) and Eq.~(\ref{deltaphisim}), the analysis result is shown in Fig.~\ref{gaymaGLSPsame}. Detailed analysis can be found in Appendix~\ref{App:B}.
    \begin{figure}[htb]
        \centering
        \includegraphics[width=0.45\textwidth]{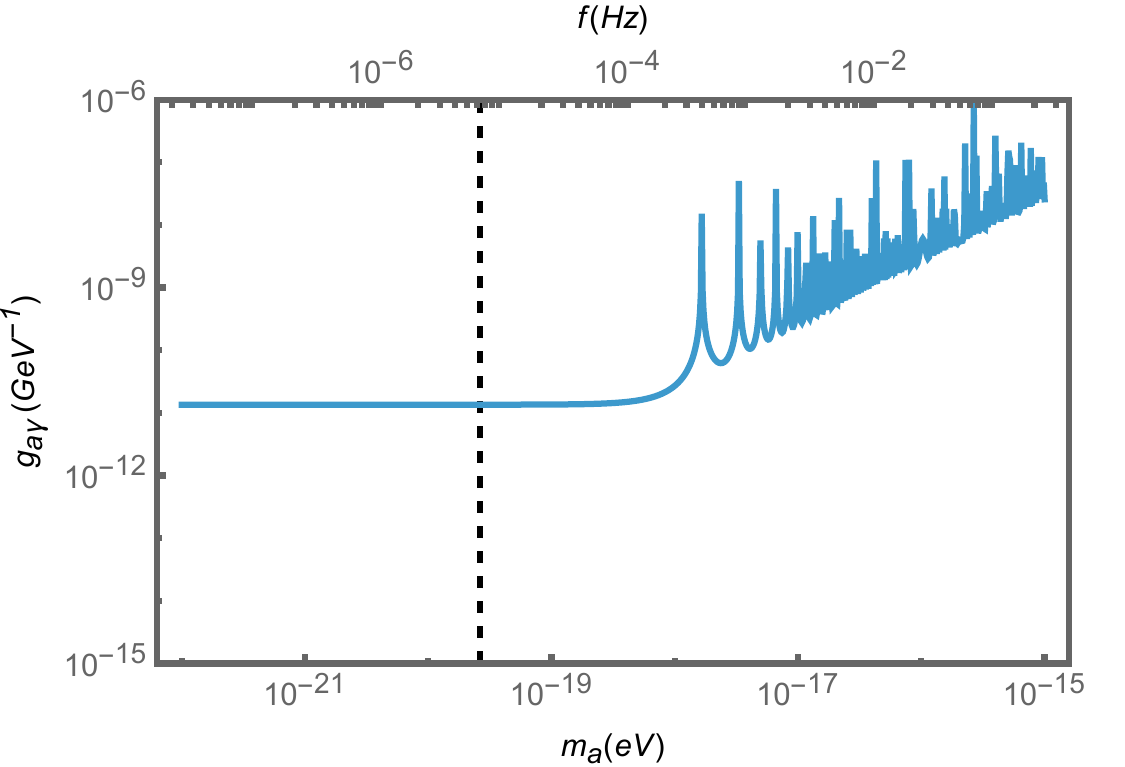}
        \caption{Sensitivity curve of the APPA satellite network with frequentist analysis. The detection sensitivity is set to $\phi_{a}=0.003^{\circ}$, with minor fluctuations attributed to instrumental noise. The black dashed line marks the critical boundary, to the left of this boundary, the condition $\mathrm{cos}\big(m_{a}T\big)\simeq1$ holds true, while to the right it is no longer valid (we adopt $m_{a}T=0.1$ as the reference criterion).}
        \label{gaymaGLSPsame}
    \end{figure}

\section{Conclusion}\label{Sec:4}
Inspired by the artificial precision timing array in \cite{2401.13668}, we investigate the APPA concept, which leverages an artificial satellite network to achieve precise measurements of ultralight scalar axion signals. Compared with PPA data obtained from ground-based observatories, APPA offers several significant advantages: (i) The number of uncertain or uncontrollable physical parameters is greatly reduced, since most parameters can be explicitly determined and flexibly adjusted; (ii) Complex or periodic physical disturbances are largely eliminated, thereby reducing the burden of data analysis and enhancing the quality of the raw observational data; (iii) Due to the ultra precise clocks of satellites and the tunability of pulsed signal parameters, the S/N, periodicity, and long term stability of the measurements are well ensured. Consequently, APPA provides an exceptionally broad detectable frequency range.

To clarify the feasibility and advantages of APPA, we employ two complementary approaches: likelihood analysis and frequentist analysis. For the likelihood analysis, we avoid prohibitively large scale computations by using the convenient expression in Eq.~(\ref{q}) from \cite{10.1140/epjc/s10052-011-1554-0,10.1103/PhysRevD.103.076018}, which allows us to efficiently derive the $g_{95\%}$ constraint. Likelihood analysis indicates that larger scales $L$ of satellite networks are more advantageous for detecting low mass axions. However, APPA exhibits identical detection limits when axions satisfy $\lambda_{\text{c}}\ge L$, for different $L$. Fig.~\ref{compare} shows that APPA exhibits a detection advantage within $m_a\sim\mathcal{O}\big(10^{-22}–10^{-18}\big)$ eV, which lies well below the detection upper bound. Compared with traditional ground-based observations, APPA achieves a tighter $g_{95\%}$ bound, highlighting its strong detection potential. For the frequentist approach, we adopt the GLSP, a standard method for detecting periodic sinusoidal signals. MC simulations demonstrate that, even under low S/N conditions, APPA produces GLSP spectra with more pronounced target signal peaks, significantly reduced FAP, and fewer interference peaks compared with conventional observations.

With the continuous advancement of ultra precise timing and observational technologies, it may become feasible in the future to deploy the APPA satellite network in space, thereby realizing ideal ``artificial pulsars" whose physical parameters can be deliberately modulated. Such a system would eliminate many of the uncertainties inherent in natural pulsar signal sources and, in turn, open an entirely new avenue for the detection of axion signals.

\section*{Acknowledgments}
We thank helpful conversion with Profs. Szabolcs Marka and Jing Ren.
This work is supported by the National Key Research and Development Program of China (Nos. 2023YFC2206200, No.2021YFC2201901), the National Natural Science Foundation of China (No.12375059), and the Project of National Astronomical Observatories, Chinese Academy of Sciences (No.E4TG6601).

\appendix
\section{Birefringence effect of electromagnetic wave}\label{App:A}
At the beginning of the main text, in Eq.~\eqref{SL}, we introduce the Lagrangian density $\mathcal{L}$ incorporating the axion contribution. In this appendix, we provide a detailed derivation of how the axion induces birefringence in electromagnetic waves. We begin by considering the equations of motion for the electromagnetic four-potential $A_{\mu}$. Because
    \begin{equation}
        \begin{aligned}
        \frac{\partial \mathcal{L}}{\partial A_{\lambda}}=\frac{\partial \mathcal{L}}{\partial F_{\kappa \lambda} }\frac{\partial F_{\kappa \lambda} }{\partial A_{\lambda}}=0,
        \end{aligned}
        \label{a1}
    \end{equation}
and
    \begin{equation}
        \begin{split}
            \frac{\partial\mathcal{L}}{\partial\big(\partial_{\kappa } A_{\lambda }\big)}&=\big[-\frac{1}{4} g^{\mu \alpha } g^{\nu \beta }\big(F_{\alpha\beta}\delta _{\mu }^{{\kappa }' }\delta _{\nu }^{{\lambda}' }+F_{\mu\nu }\delta _{\alpha }^{{\kappa }' }\delta _{\beta }^{{\lambda}' }\big)\\
            &\quad -\frac{g_{a\gamma }}{4}a\cdot\frac{1}{2}\varepsilon^{\mu \nu \rho\sigma }\big(F_{\rho\sigma}\delta _{\mu }^{{\kappa }' }\delta _{\nu }^{{\lambda}' }+F_{\mu\nu}\delta _{\rho}^{{\kappa }' }\delta _{\sigma}^{{\lambda}' }\big)\big]\\ 
            &\quad \big(\delta^{\kappa}_{{\kappa}'} \delta^{\lambda}_{{\lambda}'}-\delta^{\lambda}_{{\kappa}'} \delta^{\kappa}_{{\lambda}'}\big) \\
            &=-F^{\kappa\lambda}-g_{a\gamma }a\widetilde{F}^{{\kappa\lambda}}.
        \end{split}
        \label{a2}
    \end{equation}
Substituting Eq.~(\ref{a1}) and Eq.~(\ref{a2}) into the Euler–Lagrange equation, we obtain
    \begin{equation}
        \partial_{\mu}F^{\mu\nu }=-g_{a\gamma}\widetilde{F}^{\mu \nu}\partial_{\mu}a.
        \label{a3}
    \end{equation}
In deriving the above equation, we have used the property $\partial_{\mu}\widetilde{F}^{\mu\nu}=0$. By subsequently expanding both it and $\partial_{\mu}\widetilde{F}^{\mu\nu}=0$ in terms of temporal and spatial components, we obtain a modified system of Maxwell's equations that includes the axion induced corrections.
    \begin{equation}
        \begin{aligned}
            \nabla\cdot \vec{E} &=-g_{a\gamma }\nabla a\cdot \vec{B},\\
            \nabla\times \vec{B}&= \partial_{t} \vec{E}+g_{a\gamma}\big[\big(\partial_{t}a\big) \vec{B}+\big(\nabla a\big)\times \vec{E}\big],\\
            \nabla\cdot \vec{B} &=0,\\
            \nabla\times \vec{E}&=-\partial_{t} \vec{B}.
        \end{aligned}
        \label{a4}
    \end{equation}
We decompose the electric field into left- and right-handed circularly polarized components,
    \begin{equation}
        \vec{E}=\big(E_{R} \vec{e}_{R}+E_{L} \vec{e}_{L}\big)e^{i (\vec{k}\cdot \vec{r}-\omega t )}, 
        \label{a5}
    \end{equation}
the wave vector $\vec{k}$ is orthogonal to the polarization basis vectors $\vec{e}_{R}$ and $\vec{e}_{L}$, and satisfies
    \begin{equation}
        \begin{aligned}
            &\vec{e}_{R}=\frac{1}{\sqrt{2}}\big(\vec{e}_{x}-i\vec{e}_{y}\big),\quad \hat{k} \times \vec{e}_{R} = i\vec{e}_{R},\\
            &\vec{e}_{L}=\frac{1}{\sqrt{2}}\big(\vec{e}_{x}+i\vec{e}_{y}\big),\quad \hat{k} \times \vec{e}_{L} = -i\vec{e}_{L}.
        \end{aligned}
        \label{a6}
    \end{equation}
By combining Eqs.~(\ref{a4})–(\ref{a6}), we obtain
    \begin{equation}
        \begin{split}
            &\big(|\vec{k}|^{2}-\omega^{2}\big)\big(E_{R} \vec{e}_{R}+E_{L} \vec{e}_{L}\big)\\
            =-&g_{a\gamma}\big[\big(\partial_{t}a\big)|\vec{k}|+\big(\nabla a\cdot \hat{k}\big)\omega\big]\big(E_{R} \vec{e}_{R}-E_{L} \vec{e}_{L}\big).
        \end{split}
        \label{a7}
    \end{equation}
Since the axion effect is perturbative, we can assume $|\vec{k}|\approx\omega$, and define the small frequency shift $|\vec{k}|-\omega =\Delta \omega$. Under this approximation, the final result is obtained,
    \begin{equation}
        \Delta \omega =\pm \frac{1}{2} g_{a\gamma }\big(\partial_{t}a+\nabla a\cdot \hat{k}\big).
        \label{a8}
    \end{equation}

Following the previous procedure for the electromagnetic field, we now solve the kinetic equations for the axion. From the Euler–Lagrange equation, we have
    \begin{equation}
        \begin{split}
            \partial_{\mu }\partial^{\mu }a-m_{a}^{2}a=\frac{g_{a\gamma }}{4}F_{\mu \nu }\tilde{F}^{\mu \nu }
            =-g_{a\gamma }\vec{E}\cdot\vec{B}.
        \end{split}
        \label{a9}
    \end{equation}
Thus, the axion satisfies the Klein–Gordon equation with a source term. The general solution of this equation can be expressed as the sum of the homogeneous solution and a particular solution induced by the source:
%The solution can be expressed as
    \begin{equation}
        a\big(x\big)=a_{\text{hom}}\big(x\big)+a_{\text{par}}\big(x\big),
        \label{a10}
    \end{equation}
where $a_{\text{hom}}\big(x\big)$ represents the free oscillatory field, $a_{\text{par}}\big(x\big)$ corresponds to the forced oscillations induced by the source term. Under the Fourier transformation $\partial^{\mu }\partial_{\mu }\to -k^{\mu }k_{\mu }$, the particular solution can be expressed as
    \begin{equation}
        a\big(k\big)=\frac{g_{a\gamma }\vec{E}\cdot\vec{B}}{m_{a}^{2}+k^{\mu}k_{\mu}}.
        \label{a11}
    \end{equation}
This expression exhibits singularities on the mass shell, to regularize the on shell divergence, one can introduce an infinitesimal imaginary term, yielding
    \begin{equation}
        a\big(k\big)=\frac{g_{a\gamma }\vec{E}\cdot\vec{B}}{m_{a}^{2}+k^{\mu}k_{\mu}+i\varepsilon}.
        \label{a12}
    \end{equation}
Consequently, the solution corresponding to the source term is
    \begin{equation}
        a_{\text{par}}\big(x\big)=\int \frac{d^{4}k}{\big(2\pi\big)^{4}}e^{-ik^{\mu }x_{\mu }}\frac{g_{a\gamma }\vec{E}\cdot\vec{B}}{m^{2}_{a}+k^{\mu}k_{\mu}+i\varepsilon},
        \label{a13}
    \end{equation}
where $k^{\mu}k_{\mu}=-\omega^{2}+|\vec{k}|^{2}$. Considering the dispersion relation, the possible cases are as follows: ($i$) $|\vec{k}|^{2}\equiv \omega^{2}-m_{a}^{2}>0$, corresponding to real, propagating waves; ($ii$) $|\vec{k}|^{2}=0$, in which the wave vector lies on the axion mass shell, leading to resonance and amplification of the field; ($iii$) $|\vec{k}|^{2}<0$, corresponding to the imaginary wave vector, where the axion field is effectively screened.

For axion-like ULDM, the Compton frequency $\nu_{c}\sim \mathcal{O}\big(10^{-8}-10^{-4}\big)\mathrm{Hz}$, which lies in extremely low frequency bands generally beyond the observational range of the Electromagnetic waves.

\section{GLSP: Generalized Lomb-Scargle Periodogram}\label{App:B}
Under the APPA framework, the satellite network is confined to a small scale region of the Solar System with $L_i\approx 5$AU (Jupiter's orbit). As mentioned in the main text, for axions velocity $v\sim10^{-3}c$, the de Broglie wavelength $\lambda_{\text{de}}$ of the scalar axion field is larger than the scale of satellite networks. Therefore, the entire network can be treated as residing within a single coherent axion field, then Eq.~(\ref{phiadelta}) reduces to Eq.~(\ref{phiapart3}).
    \begin{figure}[htb]
        \centering
        \includegraphics[width=0.45\textwidth]{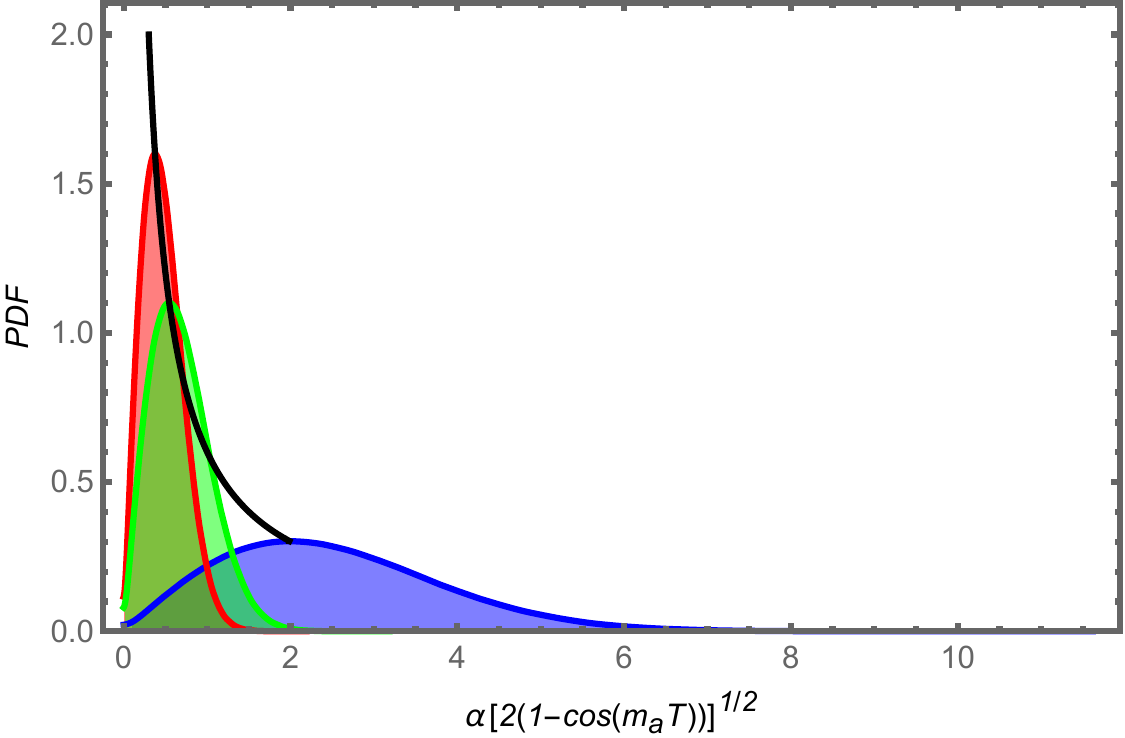}
        \caption{The Histograms and corresponding PDFs of $\alpha\big[2\left(1-\mathrm{cos}\left(m_{a}T\right)\right)\big]^{\frac{1}{2}}$ for different axion masses $m_{a}$, under the condition $\alpha_{o}=\alpha_{s}=\alpha$. The distributions were generated using the MC method with $10^{6}$ samples for each $m_{a}$. Color coding: Red, $m_{a}=10^{-19}\mathrm{eV}$; Green, $m_{a}=10^{-17.5}\mathrm{eV}$; Blue, $\mathrm{cos} \big(m_{a}T \big)=-1$. The black curve represents the envelope defined by the maxima of the PDFs across different values of $m_{a}$, which diverges at the end when approaching zero that $\mathrm{cos} \big(m_{a}T \big)=1$.}
        \label{multirldissame}
    \end{figure}

Using MC method, we derive the statistical histogram and probability density function (PDF) of Eq.~(\ref{phiapart3}) for different $m_{a}$, the results are presented in Fig.~\ref{multirldissame}. Note that for arbitrary $m_{a}$, the statistical distribution of amplitude fluctuations lies between the two limiting cases of $\mathrm{cos}\big(m_{a}T\big)=1$ and $\mathrm{cos}\big(m_{a}T\big)=-1$ (see Fig.~\ref{multirldissame}).

\begin{figure}[htb]
        \centering
        \includegraphics[width=0.45\textwidth]{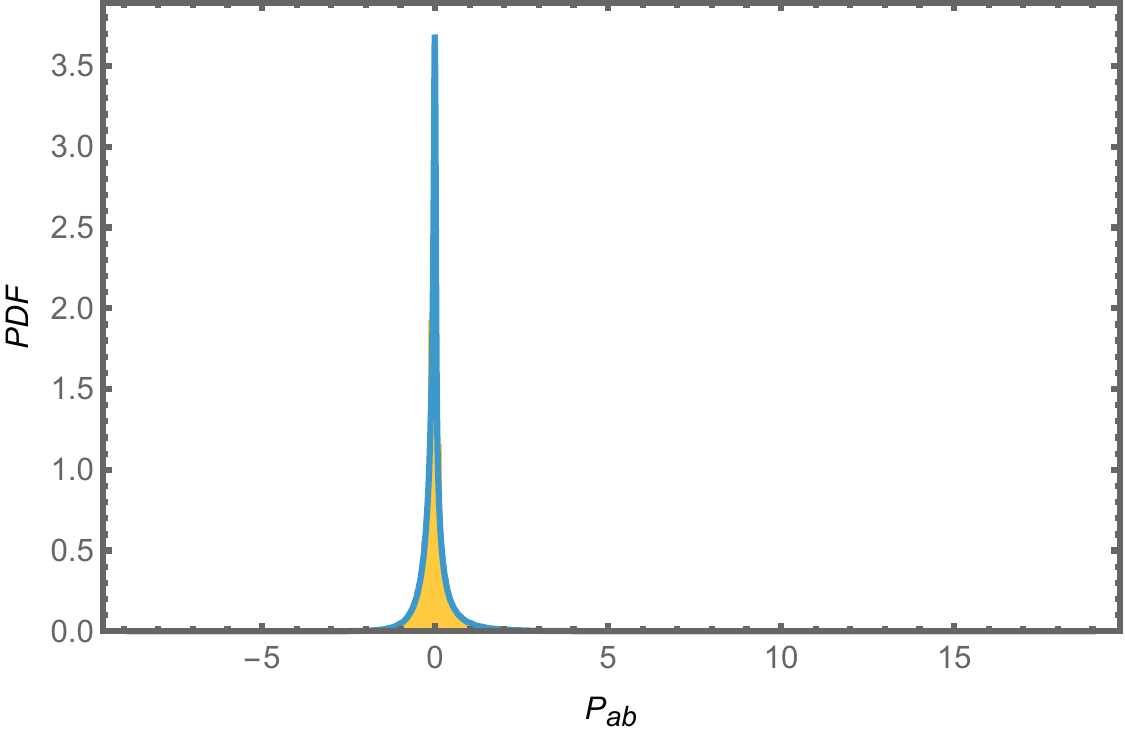}
        \caption{The histogram of Eq.~(\ref{Pabsame}) (yellow, 100 bins) together with its PDF (blue) is shown. A total of $10^{6}$ samples were generated using the MC method.}
        \label{Pabdissame}
    \end{figure}

To further simplify the analysis, we demonstrate that when processing APPA data with frequentist analysis, the cross-correlation between transmitting satellites can be safely neglected. Taking practical orbital factors into account, the distances between the transmitting satellites ($a,b=1,...,6$) and the receiving satellite are changing with time during the operation of the satellite network (as in Fig.~\ref{satellite}). According to Eq.~(\ref{phiapart3}), the signal transmission time $T_a$ and $T_b$ vary continuously, which results in the randomness of $\varphi_a$ and $\varphi_b$. Using Eqs.~\eqref{deltaphiphia} and \eqref{phiapart3}, we obtain the correlation coefficient as follows:
    \begin{equation}
        \begin{aligned}
            P_{ab}&=\overline{\Delta\phi_{a}(t)\Delta\phi_{b}(t)}\\
            &=2\left(\frac{g_{a\gamma}}{m_{a}}\sqrt{\frac{\rho_{\text{DM}}}{2}}\right)^{2}\alpha^{2}\Big\langle\big[\big(1-\cos m_{a}T_{a}\big)\big(1-\cos m_{a}T_{b}\big)\big]^{\frac{1}{2}}\\
            &\quad\times\cos\left(m_{a}t+\varphi_{a}\right)\cos\left(m_{a}t+\varphi_{b}\right)\Big\rangle\\
            &\propto \alpha^{2}\Big\langle\big[\big(1-\cos (m_{a}T_{a})\big)\big(1-\cos (m_{a}T_{b})\big)\big]^{\frac{1}{2}} \\&\quad\times
            \cos\big(\varphi_{a}-\varphi_{b}\big)\Big\rangle.
        \end{aligned}
        \label{Pabsame}
    \end{equation}
In the derivation, since the observation time $T_{\text{obs}}\sim \mathcal{O}\big(10^{1}\big)\mathrm{yr}$ is extremely long, the time averaged value of the high frequency oscillatory term $\cos\big(2m_{a}t+ \varphi_{a}+\varphi_{b}\big)$ over the observation period is effectively zero. To estimate the correlation coefficients $P_{ab}$, a total of $10^{6}$ samples were generated using the MC method, and the PDF of $P_{ab}$ is shown in Fig.~\ref{Pabdissame}. The results indicate that the correlation coefficients between different transmitting satellites are  clustered around zero. Therefore, when analyzing APPA data with GLSP method, the inter signal correlations between distinct satellites can be neglected (see e.g. \cite{10.1103/PhysRevD.111.062005}).

\bibliography{ref}

\end{document}